\documentclass[conference]{IEEEtran}

\IEEEoverridecommandlockouts
\usepackage{cite}
\usepackage{amsmath,amssymb,amsfonts}
\usepackage{algorithmic}
\usepackage{graphicx}
\usepackage{textcomp}
\usepackage{xcolor}
\usepackage{amsmath}
\usepackage{breqn}
\usepackage{amsmath,amssymb,amsfonts}
\usepackage{algorithmic}
\usepackage{graphicx}
\usepackage{textcomp}
\usepackage{xcolor}
\usepackage{algorithm}
\usepackage{algorithmic}
\def\BibTeX{{\rm B\kern-.05em{\sc i\kern-.025em b}\kern-.08em
    T\kern-.1667em\lower.7ex\hbox{E}\kern-.125emX}}
\begin{document}
\title{Resource Allocation in Cloud Computing Using Genetic Algorithm and Neural Network}
\author{\IEEEauthorblockN{1\textsuperscript{st} Mahdi Manavi}
\IEEEauthorblockA{\textit{Department of Computer Science } \\
\textit{University of Houston}\\
Houston, USA \\
Mmanavi@central.uh.edu}
\and
\IEEEauthorblockN{2\textsuperscript{nd} Yunpeng Zhang}
\IEEEauthorblockA{\textit{ Department of Information Science Technology} \\
\textit{University of Houston}\\
Houston, USA\\
yzhan226@central.uh.edu}
\and
\IEEEauthorblockN{3\textsuperscript{rd} Guoning Chen}
\IEEEauthorblockA{\textit{Department of Computer Science} \\
\textit{University of Houston}\\
Houston, USA \\
gchen22@central.uh.edu }}

\maketitle
\pagestyle{plain}

\begin{abstract}
Cloud computing is one of the most used distributed systems for data processing and data storage. 
Due to the continuous increase in the size of the data processed by cloud computing, scheduling multiple tasks to maintain efficiency while reducing idle becomes more and more challenging. 
Efficient cloud-based scheduling is also highly sought by modern transportation systems to improve their security.
In this paper, we propose a hybrid algorithm that leverages genetic algorithms and neural networks to improve scheduling. Our method classifies tasks with the Neural Network Task Classification (N2TC) and sends the selected tasks to the Genetic Algorithm Task Assignment (GATA) to allocate resources. It is fairness aware to prevent starvation and considers the execution time, response time, cost, and system efficiency.
Evaluations show that our approach outperforms the state-of-the-art method by 3.2\% at execution time, 13.3\% in costs, and 12.1\% at response time.
\end{abstract}


\section{Introduction}

Cloud computing is a model that enables demand-based network access for sharing a set of configured resources, including network, server, storage location, applications, and services while minimizing latency and reducing the need for management and interaction with the service provider~\cite{b1}. 
Cloud computing enables distributed and parallel computing~\cite{b2}, making it a common choice for big data processing that single machines with limited RAM cannot handle~\cite{b3}. 



When performing big data processing using cloud computing, consumers always wish to conduct their work in a short amount of time with the minimum cost. On the other hand, service providers aim to maximize their resource efficiency and their profits. 
One of the main challenges here is to optimize resource allocation in cloud computing. It is becoming more and more critical due to the growth of cloud computing consumers and the need to meet the computing demands of modern technology~\cite{b4}. 
Recently, the rising vehicle traffic and extensive cloud applications pose a cybersecurity challenge for vehicular networks due to limited space and computing capacity on vehicular devices \cite{cchen} \cite{lliu}. This limitation increases the network's vulnerability to potential cyber threats, such as Distributed Denial of Service (DDoS) attacks and data breaches. Applications like that demand an efficient cloud-based scheduling to enhance their security.

Methods based on greedy algorithms~\cite{junqin} and genetic algorithms~\cite{cui} have been proposed to schedule tasks for neural network applications, providing optimal solutions. However, one notable limitation of these approaches is the high execution time required by the genetic algorithm. 
To reduce the computation cost of task scheduling, we propose to combine neural networks and genetic algorithms to develop a comprehensive solution for effective resource allocation. This integrated approach proves particularly effective in addressing big data problems, optimizing both time and space utilization~\cite{Xinlei}. Our method optimizes system performance and improves resource utilization across diverse computing paradigms.

Resource allocation can indirectly affect other challenges, such as performance and load balancing. In this paper, we focus on resource allocation and scheduling. The purpose of scheduling is to assign tasks to limited resources in an appropriate way~\cite{b5}. The parameters to be considered include: 
\begin{itemize}
\item Fairness: It means that all tasks should equally use the resources, or the resources are assigned to them based on the weight given to them.
\item Optimal energy consumption: This means turning off a number of servers and hosts to reduce energy waste on cloud computing.
\item Make span: It is the shorter length of interval that causes the tasks to be accomplished sooner.
\item Load balancing: This means that tasks are allocated to resources in a way that prevents some resources from being idle while others are overloaded.
\item Cost: Total cost is acquired from cloud consumers for the services that they need. This parameter can include different parts, including the cost of processors, the cost of data storage, and the cost of transferring data in the network.
\item System efficiency: it is the maximum use of resources with the minimum amount of waste resources and time.
\end{itemize}

Among the above parameters, the fairness parameter for tasks to prevent starvation was not considered by the previous studies on resource allocation in cloud computing. 
Some studies have concentrated on a small number of parameters, while others have ignored some that have an impact on the system's overall efficiency. For example, Godhrawala et al.~\cite{god} focused on quality of service but not the cost of execution for each task.
Additionally, in numerous papers that present a combined solution using a genetic algorithm and a neural network, the parameters of the neural network are determined using the genetic algorithm~\cite{Xin}. This can lead to a dependency between the two methods, and mistakes in the genetic algorithm can result in an incorrect configuration of the neural network; consequently, the results may fall short of expectations. 



To address the above issues, our work makes the following contributions
\begin{itemize}
    \item We propose a novel scheduling method for cloud computing. Our method combines genetic algorithms with neural network techniques. Different from the previous methods, we use a neural network to select tasks to be sent to the genetic algorithm for scheduling. Our approach is customizable and can be adapted to different cloud computing environments and requirements, allowing for dynamic changes in resource allocation requirements in terms of the weight of each parameter. The approach can adapt to changing conditions in cloud computing environments, ensuring that resources are allocated optimally.
    \item Our model can be configured to consider different factors such as execution time, response time, utilization, and cost. It is also fairness aware and prevents starvation for any task to allocate resources optimally.
    \item Due to using a trained model for the classification and selection of a set of optimum tasks, we introduce a new model which improves scalability by efficiently allocating resources to meet the increasing demand for cloud computing resources. The approach can adapt to changes in the workload and allocate resources accordingly, ensuring that applications have access to the required resources.
    \item Our Neural Network Task Classification (N2TC) and Genetic Algorithm Task Assignment (GATA) can be used to gain important insights on choosing how to allocate resources. Cloud providers allocate resources more intelligently, improving overall performance and reducing costs, by studying previous data and forecasting future resource requests using our methodology.
\end{itemize}

Compared to the state-of-the-art methods, our approach leads to 3.2\% reduction in execution time, 13.3\% reduction in cost, and 12.1\% improvement in response time.

\section{Related Work}
In this section, we briefly review the works that are closely related to the proposed method. We have classified these related works into two categories: metaheuristic-based resource scheduling and dynamic resource allocation.
\subsection{Metaheuristic-based resource scheduling}
Alkayal et al.~\cite{b6} proposed a Particle Swarm Optimization (PSO) algorithm to optimize cloud computing resource scheduling for increased efficiency. The system prioritizes tasks based on length and assigns them to virtual machines that are mapped to physical machines in the data center. 
Mezmaza et al.~\cite{b7} used the parallel hybrid genetic algorithm to find the optimal set. They used the island model to migrate the tasks. Their method is energy aware and reduces the makespan parameter. Their cloud model is implemented in a data center that is composed of heterogeneous machines. Their model has also been implemented using ParadisEO. 
Mocanu et al.~\cite{b8} proposed a genetic algorithm that uses the roulette wheel to select chromosomes.  That method uses elitism in choosing chromosomes and considers a threshold level of 20 to create generations. The goal is to minimize the execution time. The fitness function focused on utilization. It is computed by dividing the total assigned input sizes by the max span. 
Geetha et al.~\cite{b11} proposed an integrated neural network and genetic algorithm for scheduling. They  reduced the context switching in the processor to save energy. Their approach handles unlimited requests in a parallel and distributed system. They also focused on a federated cloud. 
Zhou et al.~\cite{zhou} presented a Growable Genetic Algorithm (GGA) using a Heuristic-based Local Search Algorithm (HLSA) and a Random multi-Weight-based algorithm. Their method introduces a growth stage to the genetic algorithm, resulting in GGA, which allows individuals to evolve through different growth routes.
Ajak et al.~\cite{ajak} introduced a Directed Acyclic Graph(DAG) scheduling model aimed at optimizing the quality of service parameters in the cloud computing  platform. Their primary objective is to achieve makespan optimization through the appropriate allocation of tasks to nodes and arranging the execution sequence of jobs/tasks. 
To achieve near-optimal solutions, the proposed model leverages resource provisioning and heuristic techniques. 

\subsection{Dynamic resource allocation}
Praveenchandar et al.~\cite{b9} proposed a dynamic resource allocation that is energy aware and considers the size of a task and an inter arrival time. 
The method can improve response time, resource utilization, task completion ratio, and makespan. It also improves the efficiency of the dynamic resource allocation process. The authors used Cloudsim to simulate the method and they compared the model with first come first served and round-robin. 
In the model, they used a dynamic resource table updating method.
Semmoud el al.~\cite{b10} introduced a technique to achieve load balancing on their network and minimize idle time and make span. The authors limited task migrations when the load of VMs is greater than the starvation threshold and used task priority level for the quality of service in cloud computing. They used Cloudsim for the simulation. In the simulation part, sixteen data centers are considered which are located in different regions, and each of them has five physical machines. 
Shin et al.~\cite{b12} proposed a multiple adaptive resource allocation with a real-time supervisor scheme. They used hybrid cloud services for the industrial internet of things to implement their model. 
To improve response time and reduce cost, they provided the optimal number of virtual machines. In addition, they used karush-kuhn-tucker optimization that is applied to continuous time Markov chain, and all resources in public and private cloud are fully considered.

In~\cite{b6}\cite{b7}\cite{b8}\cite{b11}\cite{zhou}\cite{ajak}\cite{b9}\cite{b12}, fairness is not considered, thus, there is a possibility of starvation for tasks. Moreover, due to the use of penalty functions in~\cite{ajak}\cite{b10}\cite{b12}, the computation overhead can be high. More importantly, some approaches pay little attention to balancing critical parameters in resource allocation, leading to sub-optimal scheduling. Our method aims to address these issues.

\section{Proposed Approach}

In this section, we provide a detailed description of our approach, including the architecture and the involved algorithms of our approach.
We used a hybrid cloud to implement our model which combines both public and private cloud services.

\subsection{Architecture}

Task scheduling is essential in cloud computing. Since the cloud provider has to deal with many user applications, task scheduling can no longer be handled by traditional schedulers~\cite{b13}. Figure \ref{fig1} illustrates the architecture of our proposed system, which consists of 5 components:
\begin{figure}[!h]
\centerline{\includegraphics[width=1.\columnwidth]{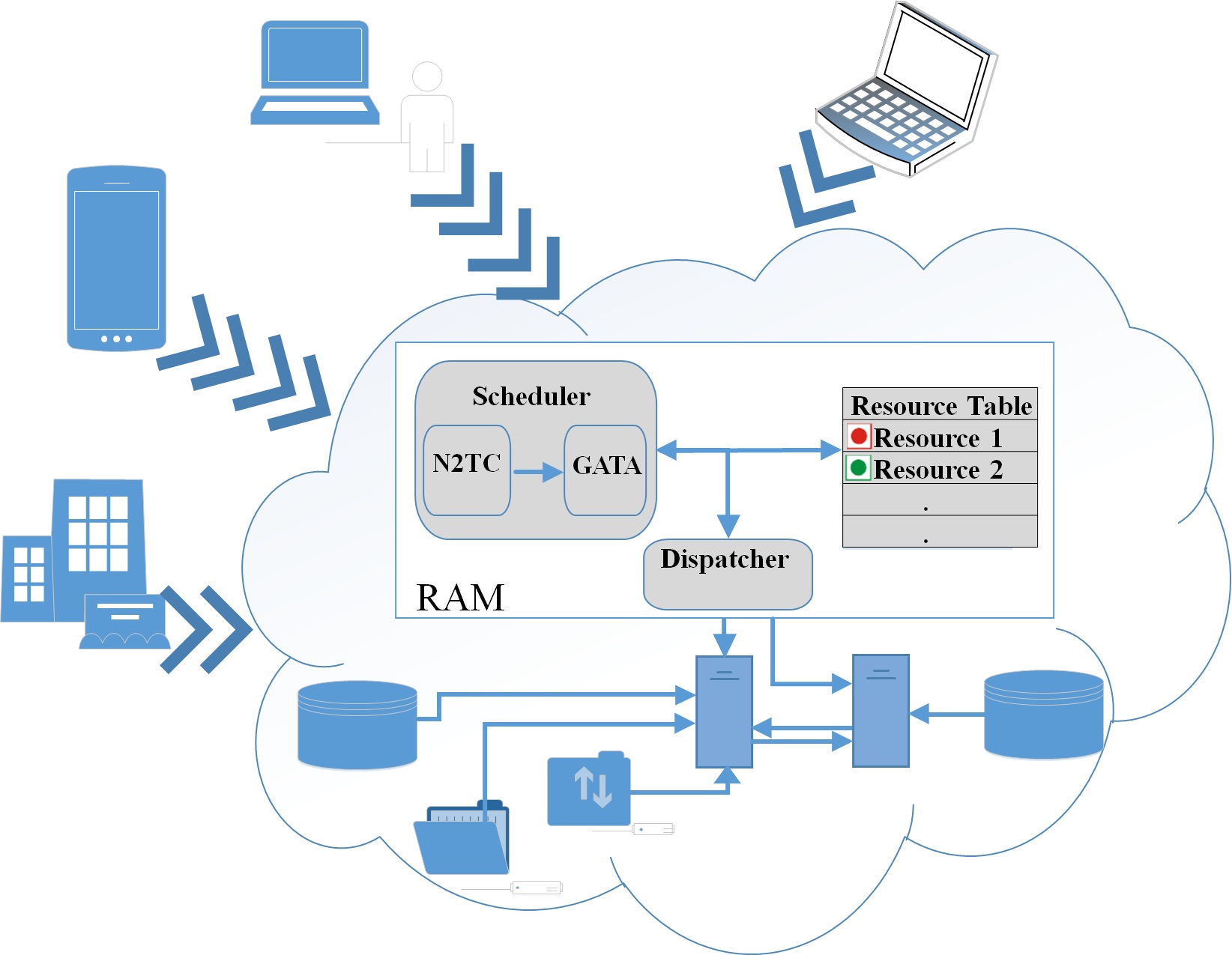}}
\caption{High-level architecture consisting of a scheduler, clients, dispatcher, resource table, and resources.}
\label{fig1}
\end{figure}

\begin{itemize}
\item Scheduler: It consists of 2 components, Neural Network Task Classification (N2TC) and Genetic Algorithm Task Assignment (GATA), which is the main module for scheduling and all computation processes are performed on it.
\item Dispatcher: It sends the tasks to the resources based on the situation of resources and binds the tasks to resources.
\item Resources: All tools and applications in server-side that are used to respond clients requests, such as fetch a file or computation request.
\item Resource table: It shows the current status of resources and the number of tasks that each resource is running right now. The resource table is necessary so that we can keep track of the resources' status and use any that are not in use during the current scheduling period.
\item Clients: An entity that wants to use cloud computing.
\end{itemize}

In our architecture, at first, tasks are sent by clients to the RAM. We use online mode to send a task.  Depending on which resources the task requires, RAM sends a request to the resource table module to determine the status of the intended resources. According to the number of available resources, the appropriate tasks are selected by GATA and N2TC, then transmitted to the dispatcher along with the free resource specification. Dispatcher then transfers the received tasks to the resources. The role of the dispatcher is to send tasks to the desired resources and to ensure those tasks are received by the resources (e.g., via the acknowledgments of resources). Next, the status of resources in the resource table is updated, and it waits for the next task to be assigned by the schedule. The resource table is intended to monitor the resources status of the network. When the scheduler is aware of the current state of resources on the network, it can perform scheduling more efficiently. Also, the resources in this architecture have a number of virtual machines to increase the speed of task execution. The proposed architecture increases the accuracy of selecting the appropriate tasks as well as the convergence towards the optimal tasks by classifying the tasks and then selecting the optimal set that is performed in the  scheduler.

\subsection{Mathematical Model}

In this section, we present the definitions, methods, and equations used in N2TC and GATA that are the essential steps of RAM module shown in Figure \ref{fig1}. 
In particular, we provide the necessary formulas for weighing the tasks and their classification based on the desired parameters and a fitness measurement. 
Table \ref{tab1} lists the abbreviations used in this paper.

\begin{table}[htbp]
\caption{Table of abbreviations}
\begin{center}
\begin{tabular}{|c|c|}
\hline
\textbf{abbreviations  }&{\textbf{Meaning}} \\
\cline{1-2} 
TW & Task Weight\\
\hline
ET & Execution Time\\\hline
C & Cost\\\hline
SE & System Efficiency\\\hline
WP & Weight of Parameter\\\hline
CW & Class Weight\\\hline
TC & Tasks of Class \\\hline
SC & Size of Class\\\hline
RT & Response Time\\\hline
F & Fairness\\\hline
Q & Queue\\\hline
r & Class Number (1,3)\\\hline
MR & Minimum Resources required\\\hline
\end{tabular}
\label{tab1}
\end{center}
\end{table}

Equation (\ref{eq:weighted}) computes the weight for a task $i$ based on the parameters of the execution time, cost, and system efficiency. We assign the initial weight to each parameter based on our system condition dynamically.
We estimate the parameter values for a new task that is added to the network, using limited historical data~\cite{shah}.
\begin{dmath}
\scalebox{0.77}{$
TW_{[i]} =[WP(ET) \times ET _{[i]} ]+[WP(C) \times C_{[i]} ]+[(WP(SE) \times SE_{[i]} ]$}
\label{eq:weighted}
\end{dmath}
\noindent where WP(ET) is the weight determined for execution time. WP(C) is the weight of cost, and WP(SE) is the weight of system efficiency. $ET_{[i]}$ is the execution time, $C_{[i]}$ is the cost, and $SE_{[i]}$ is the system efficiency of task $i$.

In Equation (\ref{eq:average}), the weight of a task is compared with the average weights of different classes of tasks. The tasks with similar weights are grouped into a class. $TW_{[i]}$  is the total weight of task $i$ (Equation (\ref{eq:weighted})), $CW_{[r]}$ is the average weight of class r, and $TC_{[r]}$ represents the tasks in the class r.
\begin{dmath}
\scalebox{1}{$
\forall i\varepsilon[1..n],| TW_{[i]} - CW_{[r]} |<\epsilon\rightarrow TC_{[r]}=i\bigcup TC_{[r]} $}
\label{eq:average}
\end{dmath}

Equation (\ref{eq:fitness}) determines the value of the fitness for each task based on response time and cost. The goal is to minimize the value of the fitness function. $SC_{[r]}$  is the size of class r, q is the class number, and p is the task number. $TC_{[q,p]}$ denotes task p in class q. 
F is the parameter of Fairness.
\begin{dmath}
Min[
\sum_{q=1}^{r}\sum_{p=1}^{SC{[r]}} RT(TC_{[q,p]}+C(TC_{[q,p]}) \times F ]
\label{eq:fitness}
\end{dmath}

If a task $i$ is waiting in the queue $Q$ from the previous scheduling, using Equation (\ref{eq:fair}) the value of the fitness function will be improved (i.e., multiplied by a factor of 0.9 to reduce the fitness value). This method is used for increasing the chance of tasks that are waiting in the queue to be executed in the next iteration.
\begin{dmath}
\scalebox{1}{
$\begin{cases}
  F_i=0.9 & \text{for }i \varepsilon Q\\    
  F_i=1 & \text{otherwise}    
\end{cases}
$
}
\label{eq:fair}
\end{dmath}

\subsection{Proposed Scheduling Algorithm}

Algorithm 1 describes the process of the proposed scheduling given all the input tasks.
First, based on the weight attributes of each task, it is added to the desired class. We implement 3 classes. In the next step, if the task has been waiting in the queue from the previous scheduling, we improve the rank of the class that the task is classified to. For example, if the class of the task is 2 and it has been waiting in the queue from the previous step, the task is transferred to class 1. When all tasks are placed in the appropriate classes, the number of idle resources in the network is compared with the number of tasks in class 1. If the number of idle resources is less than or equal to the number of tasks in class 1, then tasks of class 1 are sent to the genetic algorithm; otherwise, the tasks of class 2 are also sent to the genetic algorithm. If the idle resources are still available after sending all tasks in classes 1 and 2, we move forward to sending the tasks in class 3.
The amount of idle resources is important for us, as resources are limited. Depending on how many tasks are active on the resources at any given time, there may be a different amount of idle resources throughout execution.

\begin{algorithm}[t]
\caption{Proposed Algorithm}
\label{Proposed Algorithm}
\footnotesize
\begin{algorithmic}[1]
\renewcommand{\algorithmicrequire}{\textbf{Input:}}
\renewcommand{\algorithmicensure}{\textbf{Output:}}
\REQUIRE List of tasks
\ENSURE Optimum set of tasks
 \FOR {$i =1$ to $n$}
 \FOR {$j = 1$ to $3$}
\STATE \IF {(Task i is similar to a set of class j)}
\STATE \IF{(Task i in waiting queue and $j>$1)}
\STATE Add task i to class(j-1)
\ELSE
\STATE Add task i to class j
\ENDIF

\ENDIF
 \ENDFOR
\ENDFOR
\WHILE{($Number of Idle Resources>$set of tasks)}
\STATE Add jobs of class 2 or class 3 to a set of tasks for scheduling

\ENDWHILE
\STATE Initial population(set of tasks)
\IF{(Each gene in waiting queue)}
\STATE Fitness of gene improve 0.1
\ENDIF
\STATE Sort chromosome by DESC
\WHILE{((!Feasible solution) OR (Iteration !=Max))}
\STATE Select parents by Elitism
\STATE Apply two-point crossover
\STATE Gene of a chromosome is muted
\STATE Local search in the gene muted for finding a better gene to replace
\STATE Apply mutation
\ENDWHILE
\RETURN $Set of Tasks$ 
\end{algorithmic} 
\end{algorithm}

Next, the selected tasks are fed to the genetic algorithm. The initial population is constructed randomly, and the value of the fitness function for each chromosome is calculated. If there is any task waiting in the queue from the previous scheduling, the value of the fitness function of that gene will be increased by 10\%. Then the chromosomes are sorted according to their fitness values, and the higher chromosomes are used for the offspring generation. We use the two-point method, and the intersection operator is applied to them. To apply the mutation operator, the mutation is initially performed on the desired gene. Then, a local search is performed around the mutated gene so that if there is a gene with a higher value of the fitness function, it will be selected. This process is performed until the optimal set is found or the iteration number of the process exceeds the predetermined number of iterations.
The time complexity analysis of our algorithm is as follows.
The first nested loop contributes $O(n)$ to the overall time complexity, where $n$ is the number of tasks.
The while loop depends on the number of idle resources $m$, and its time complexity is $O(m)$.
The initial population setup has a time complexity of $O(n)$.
The sorting process takes $O(n \log n)$ time.
The last while loop has a time complexity of $O(kn)$, where $k$ represents the number of iterations until a feasible solution is found. The overall time complexity is $O(n \log n + kn + m)$. If $k$ and $m$ are much smaller than $n$, the overall time complexity simplifies to $O(n \log n)$.

\subsection{N2TC}

N2TC is used to classify the input tasks that are entered into the cloud computing, which operates on the basis of the neural network.
We use a feed-forward back propagation neural network for our model.

\subsubsection{Data Preparation}

In the first step, data are partitioned. 70\% of the data is used for training, 15\% for network validation, and 15\% for network testing. Although there is no set ratio, 70:30 is typically regarded as the norm \cite{racz}.
The data for training, validation, and testing are selected randomly so that the performance of training, validation, and testing is enhanced.

\subsubsection{Transform}

Since the sigmoid logarithmic transfer function is a derivative function, it is commonly used in multi-layer networks that are trained by the back propagation algorithm. An example of this function is based on Equation (\ref{eq:frc}):
\begin{equation}
A= \frac {1}{1+e^{-net} }~ where,~ net= \sum_{i=1}^{n}(W_{[i]} * X_{[i]} )
\label{eq:frc} 
\end{equation}
$X_{[i]}$ is the input of the function and $W_{[i]}$ is the weight of the $X_{[i]}$.

\begin{figure}[h]
 \centering
 \includegraphics[width=0.6\linewidth,height=4cm]{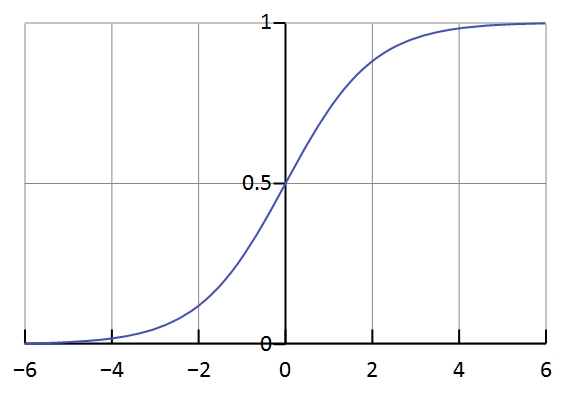}
 \caption{This illustrates the function (Equation (\ref{eq:frc})) for transform.}
 \label{fig:2}
 \end{figure}
 
\subsubsection{Training}

The scaled conjugate gradient method is used to update the weight and bias of the data, and training is terminated if:
\begin{itemize}
\item Maximum number of epochs has been created
\item The time exceeds the maximum level
\item The network's performance falls below a threshold
\item The gradient of the performance graph is below the minimum
\item Validity confirmation performance has decreased since the last time.
\end{itemize}

\subsubsection{Performance Function}
In Equation (\ref{eq:prf}), the efficiency function is applied. This function calculates the average squares of the errors between the output and the target.
One of the cases of stopping earlier before the completion of the network training is the reduced network performance. If the difference between the output and target increases, the network training will be stopped.
\begin{equation}
Performance= \frac {1}{n} \sum_{i=1}^{n} (Y^*(i) - Y(i))^2 
\label{eq:prf} 
\end{equation}
Another feature considered for this network is memory reduction, which speeds up network execution. The higher number of these layers enhances the network accuracy but increases the run-time. The number of hidden layers of the neural network is set as 20 which produces the best results based on our evaluation. 

The tasks in N2TC are divided into three categories and the high-priority tasks are sent to the GATA.
The task classification criteria in N2TC include the execution time, cost, and system efficiency. The tasks from the previous periods in the waiting queue are improved by one level to give them an opportunity to run and prevent starvation for the tasks. Also, if the number of tasks with the first priority is lower than the resources available in the cloud computing, tasks with a lower priority will be sent to GATA to have the maximum resources available on the network.

\subsection{GATA}
GATA is designed to select tasks using the cloud computing resources based on the genetic algorithm. 
In our proposed approach the method of decimal is used to represent the chromosomes because the binary display method increases the amount of data storage. All resources have a decimal number and each gene in the chromosome stores this number.

\subsubsection{Initial Population}
The first stage of the genetic algorithm is the production of the initial population. To prevent early convergence, the initial population is randomly selected to cover a wide range of data. The fitness of the chromosomes is based on their gene fitness; the initial population is 500. 
\subsubsection{Selection Function}
In this operator, from the chromosomes in a population, a number of chromosomes are selected for reproduction. The elitism method is used to select the parent chromosomes to produce the children so that the chromosomes are originally arranged on the basis of fitness value and then the chromosomes with the highest fitness value are prepared for the child generation stage. This method increases the convergence rate to achieve the optimal response.
\subsubsection{Crossover}
As part of the integration process, parts of the chromosomes are replaced randomly. This makes the children have a combination of their parents' characteristics while do not exactly resemble their parents. In our model, we use the 2-point crossover approach, and various parts of the parent chromosomes are selected for the production of children.
\subsubsection{Mutation}
After completing the crossover, the mutation operator is performed. This operator randomly selects a gene from the chromosome and changes the content of that gene. The mutation operator is used to avoid getting stuck in a local maximum or minimum. The probability of mutation in our model is 5\%.
\subsubsection{Fitness}
To solve the problem using the genetic algorithm, an appropriate fitness function must be developed for that problem. If an appropriate function is selected, higher convergence is obtained, the algorithm runs faster and the optimal answer is selected. As seen in Equation (7), we considered the response time and cost in the fitness function.
\begin{dmath}
\scalebox{0.9}{
$
fitness=\begin{cases}
  \sum_{i=1}^{n} (RT(i)+MR(i))&\text{$Task_{[i]}$ $\notin$ Q}\\  \\  
  0.9 \times [\sum_{i=1}^{n}( RT(i)+MR(i))]  & \text{otherwise}
\end{cases}
$
}
\label{eq:fair2}
\end{dmath}
In the fitness function, the minimum level of fitness value represents the optimality of the chromosome. MR(i) is the number of resources required for the task(i). Since the number of resources needed to do it is lower, it is more desirable in terms of cost. RT(i) indicates the response time for task(i) which should be minimized. After determining the fitness function for each gene, the fairness parameter is raised by asking whether task(i) has remained in the queue from the previous scheduling. If it is in the queue, its fitness value will be increased by 10\% to give it an opportunity to obtain resources to prevent starvation.

\section{EVALUATION}
We use Google cluster-traces v3 dataset\cite{wil} for evaluation. The Google dataset concentrates on resource requests and usage, without any information about end users and their data or storage systems, and so on. It consists of 405894 rows.

Table 2 shows the hardware system used to run the proposed approach.
Three assumptions are considered for the proposed approach, including:
\begin{itemize}
\item The tasks are independent from each other--i.e., to do the task i there is no need to do the task j before it.
\item Tasks do not have a deadline.
\item Tasks in our network are non-pre-emptible, the resources will not release until the task is completed. 
\end{itemize}

\begin{table}[!h]
\caption{Platform properties}
\begin{center}
\begin{tabular}{|c|c|}
\hline
\textbf{Properties  }&{\textbf{Values}} \\
\cline{1-2} 
Computer & Asus\\
\hline
CPU & Intel® Core™ i5-3230M CPU @ 2.60 GHz \\\hline
RAM	 & 6.00 GB(5.45 GB Usable)\\\hline
Operating System & 64-bit , windows 8\\\hline

\end{tabular}
\label{tab2}
\end{center}
\end{table}

We use MATLAB to implement and evaluate our proposed approach.
All tasks have been entered into the network and are awaiting scheduling. During the scheduling, we choose 10 tasks for execution based on the specified parameters. In our model, the evaluation is conducted using a set of 10 tasks, which we have determined to be sufficient for assessing the performance and effectiveness of our approach.
In our genetic algorithm, each gene in the chromosome corresponds to a specific task. It is common for genetic algorithms to utilize a relatively small number of genes in each chromosome, as observed in studies such as \cite{jia} and \cite{saleem}, where the size of the chromosomes typically ranges from 8 to 12 genes. However, it is important to note that as the size of the chromosome increases, the computational complexity of the algorithm also grows \cite{waqar}. This poses a significant constraint, as the computational demands escalate with larger chromosomes.
The response time, execution time, cost, and performance of particular tasks vary. In this section, we demonstrate that the set of tasks chosen by our model is superior to the set chosen by other approaches.
As we mentioned earlier, N2TC receives all tasks and classifies them based on metrics for execution time, cost, and system efficiency.

\begin{figure}[!t]
\center
  \includegraphics[height=5.4cm, width=\linewidth]{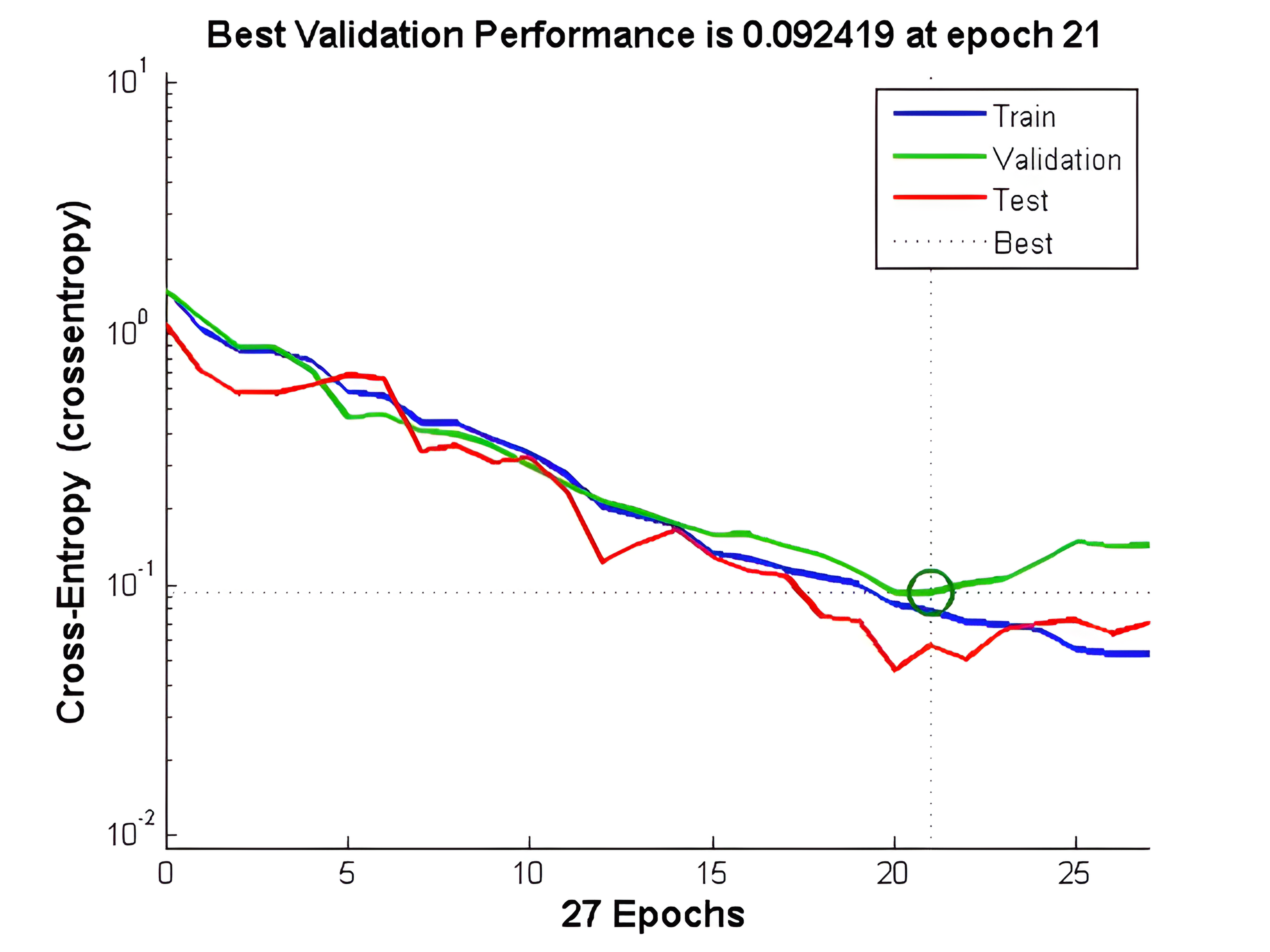}
\caption{Performance of the network during the training, validation, and testing processes. We use cross-Entropy to check the condition of our model in each epoch. In our model, we have 27 epochs.}
\label{fig:3}       
\end{figure}

Figure \ref{fig:3} presents the network performance graph.
In Figure \ref{fig:3}, the network performance is finished with 27 epochs, and it has the best performance in epoch 21. As the curvature of the test graph is higher, the probability of over-fitting in the network is greater. The descending trend of the graph indicates good network performance. Here the results of GATA are addressed. The most important part of the genetic algorithm is the fitness function.

\begin{figure}[h!]
\center
\includegraphics[width=\linewidth,height= 50mm]{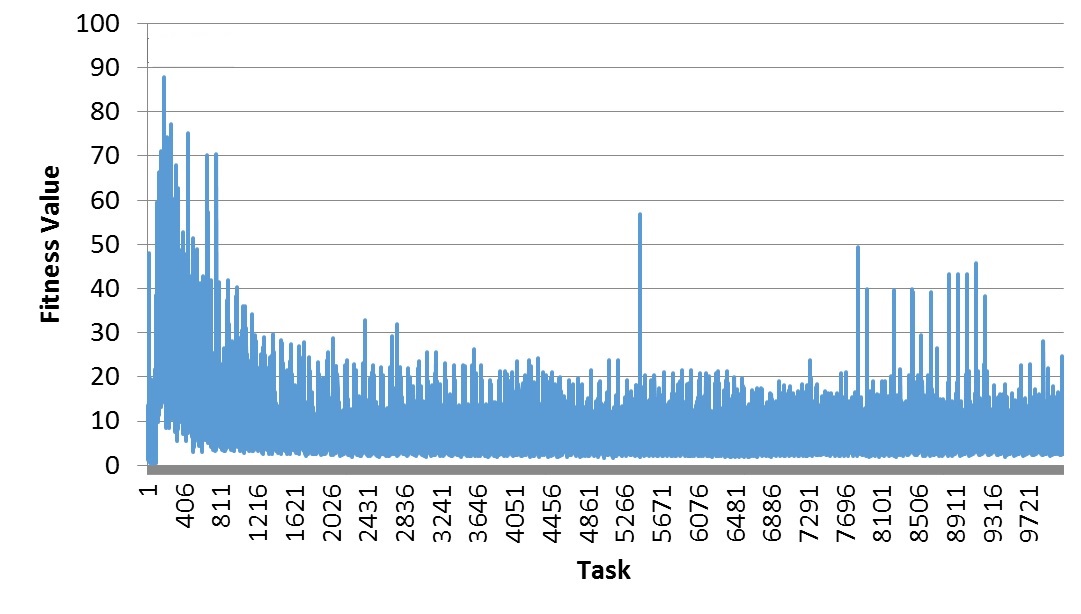}
\caption{The fitness value of selection tasks in each generation.}
\label{fig:4}      
\end{figure}

In the next step, tasks classified in class 1, in some situations tasks classified in class 2, are sent to the GATA to find an optimal set of tasks for execution. The length of the optimal set is 10, which means 10 tasks are selected for execution.\\
 Figure \ref{fig:4} presents the value of the fitness function during finding the optimal set of tasks.
 
Due to the fact that a set of tasks is optimal when the fitness value of that chromosome is minimized, the descending trend of the graph indicates the suitability of the GATA configuration.

Next, We compare the results of the proposed approach with two widely used algorithms in this field, namely First In First Out (FIFO) and Shortest Job First (SJF), as well as two related works (Mezmaz et al.~\cite{b7} and Mocanu et al.~\cite{b8}) that are considered as the best methods in this domain. These related works provide valuable insights and serve as benchmarks for comparison, as they have achieved significant advances in resource allocation techniques.

\begin{figure}[!h]
\center
\includegraphics[width=\linewidth,height= 55mm]{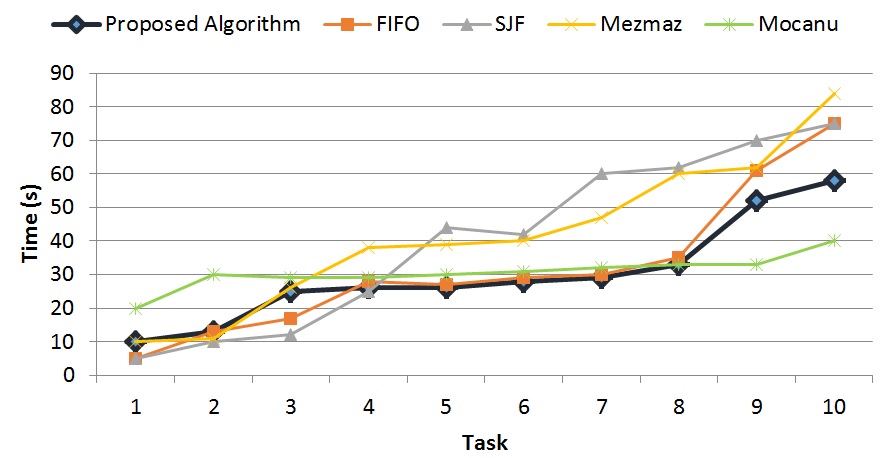}
\caption{The execution times of the 10 tasks scheduled with five different methods.}
\label{fig:5}       
\end{figure}

In Figure \ref{fig:5}, the execution times of the tasks 
with the five methods are shown, respectively.
Among the 10 selected tasks to run in each of the five methods, the proposed solution has a lower execution time in general and SJF has the longest execution time.

\begin{figure}[!h]
\center
\includegraphics[width=\linewidth,height= 52mm]{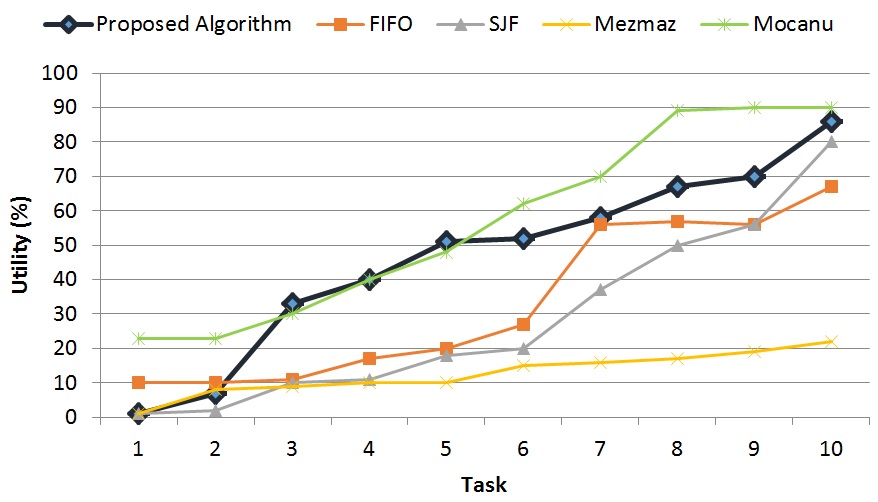}
\caption{System utilization rate with the five scheduling methods.}
\label{fig:6}       
\end{figure}

The system utilization rate is shown in Figure \ref{fig:6}. An ideal utilization rate is achieved if the available resources on the network are used maximally, and the idle resources are minimized. In other words, the higher the utilization rate, the better the scheduling is. At this point, the solution provided by Mezmaz et al. ~\cite{b7} has the worst utilization rate, while our method has the second-best utilization rate.

\begin{figure}[!h]
\center
\includegraphics[width=\linewidth, height= 52mm]{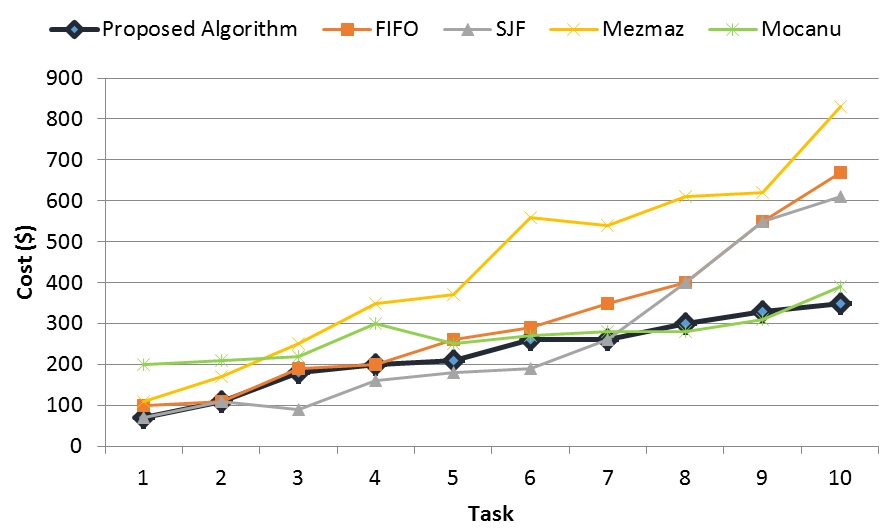}
\caption{Costs of executing the 10 tasks with five different methods.}
\label{fig:7}       
\end{figure}

Figure \ref{fig:7} compares the costs of executing the tasks with the five methods, respectively.
We see that the lowest cost of completing all 10 tasks (i.e., adding all costs for the 10 individual tasks) is with the proposed approach.

\begin{figure}[!h]
\center
\includegraphics[width=\linewidth,height= 52mm]{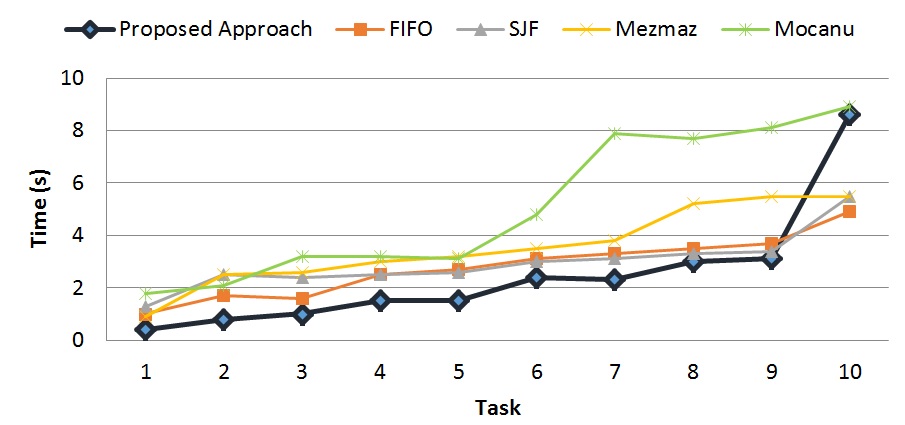}
\caption{Response times of executing the 10 tasks with the five scheduling methods.}
\label{fig:8}       
\end{figure}

Figure \ref{fig:8} shows the response time graph of the 5 scheduling methods. This measure shows the time interval between sending the task to the cloud computing and receiving the first response from the network to the user.
From the comparison, we see that our method has the shortest response time for most tasks except for task 10.

Generally, according to the presented graphs, it can be concluded that the proposed approach has the best performance among the five solutions, and it can be used in a wide range of applications.\\
Table 3 presents a summary of the performance of the stated strategies, which indicates the total average of the ten selected tasks in each solution. We converted the value of each task between the 0 and 1 range to simplify the results. Compared to the average performance of the existing state-of-the-art methods, the proposed method has improved by about 3.2\% at execution time, 13.3\% in costs, and 12.1\% at response time. 
\begin{table}[h]
\caption{Summary of the qualitative comparison between the proposed algorithm and state-of-the-art solutions. All times reported here are in seconds.}
\label{tab:summary}       
\begin {center}
\scriptsize
\begin{tabular}{|c|c|c|c|c|}
\hline
\textbf{Algorithm}& \textbf{Utility}&\textbf{Response Time}&\textbf{Cost}&\textbf{Execution Time} \\
\hline
\cline{1-5}
 Proposed Approach&0.465&\hspace{0.6cm} \textbf {0.246}& \textbf {0.227}&\hspace{0.4cm} \textbf{ 0.30}\\
\hline
 FIFO&0.331&\hspace{0.6cm}0.280&0.311&\hspace{0.6cm}0.36\\
\hline
 SJF&0.285&\hspace{0.6cm}0.296&0.262&\hspace{0.6cm}0.45\\
\hline
 Mezmaza et al.~\cite{b7}~ &0.127&\hspace{0.6cm}0.357&0.441&\hspace{0.6cm}0.41\\
\hline
 Mocanu et al.~\cite{b8}~& \textbf {0.565}&\hspace{0.6cm}0.508&0.302&\hspace{0.6cm}0.31\\
\hline
\end{tabular}
\end{center}
\end{table}

In particular, the proposed method has the best response time for nine out of ten tasks among all methods and the best costs for executing all 10 tasks. Our method also has the second-best performance in utilization rate and the best overall performance to improve the execution time.
The results from the graphs indicate that the proposed model not only prevents task starvation but also has a positive impact on the best task selection and the improvement of the aforementioned parameters. For this reason, the proposed approach outperforms the methods already mentioned.

\section{CONCLUSION AND FUTURE WORK}

Resource allocation is considered one of the major challenges in cloud computing. Many efforts have been made in this field. Since the heuristic methods have better results in large environments, these methods are more popular. 
In the proposed approach we used the combination of genetic algorithms and neural networks to solve the problem of scheduling and selection of resources to get the optimal answers for resource assignment in cloud computing. In the future, the load balancing issue can be included in the defined parameters. Also, for tasks that have a deadline, some priorities will be considered so that they can be run at the right time. In addition, this method can be extended to tasks that are dependent on each other. 
Furthermore, the proposed efficient cloud-based scheduling can be applied by transportation systems to improve cybersecurity, which enables immediate responses to potential threats and consolidates security monitoring. By harnessing the flexibility and effectiveness of cloud computing, transportation systems can strengthen their cybersecurity measures and enhance their ability to withstand cyber threats. We will explore this in the future.
\appendix[ACKNOWLEDGMENT]
This work is funded by the US Department of Transportation (USDOT) Tier-1 University Transportation Center (UTC) Transportation Cybersecurity Center for Advanced Research and Education (CYBER-CARE). (Grant No. 69A3552348332), and Theorizing Connected Vehicle (CV) Based Advanced Traffic Management System (ATMS) Vulnerability Analysis and Strategizing for Cyber Security (Grant No. I0509667)

\end{document}